\documentclass[12pt]{article}
\usepackage{graphicx}

\newcommand\pubnumber{DPF2013-233}
\newcommand\pubdate{\today}

\def\efi{Enrico Fermi Institute, University of Chicago \\
5620 South Ellis Avenue, Chicago, IL 60637}
\def\support{\footnote{Work supported in part by the U.S. Department of Energy 
          under Grant No.\ DE-FG02-90ER40560.}}

\def\Title#1{\begin{center} {\Large #1 } \end{center}}
\def\Author#1{\begin{center}{ \sc #1} \end{center}}
\def\Address#1{\begin{center}{ \it #1} \end{center}}

\newcommand\pubblock{\rightline{\begin{tabular}{l} \pubnumber\\
         \pubdate  \end{tabular}}}
\newcommand\efiblock{\rightline{\begin{tabular}{l} EFI 13-26 \qquad \qquad
          \quad \end{tabular}}}
\newenvironment{Abstract}{\begin{quotation}  }{\end{quotation}}
\newenvironment{Presented}{\begin{quotation} \begin{center} 
             PRESENTED AT\end{center}\bigskip 
      \begin{center}\begin{large}}{\end{large}\end{center} \end{quotation}}
\def\Acknowledgments{\bigskip  \bigskip \begin{center} \begin{large}
             \bf ACKNOWLEDGMENTS \end{large}\end{center}}

\def\beq{\begin{equation}}
\def\eeq#1{\label{#1}\end{equation}}
\def\eeqn{\end{equation}}

\def\beqa{\begin{eqnarray}}
\def\eeqa#1{\label{#1}\end{eqnarray}}
\def\eeqan{\end{eqnarray}}

\let\bar=\overbar

\def\Dslash{\not{\hbox{\kern-4pt $D$}}}
\def\dslash{\not{\hbox{\kern-2pt $\del$}}}

\def\msb{{\bar{\ssstyle M \kern -1pt S}}}

\begin{document}
\begin{titlepage}
\pubblock
\efiblock

\vfill
\Title{Theoretical Issues in Flavor Physics}
\vfill
\Author{Jonathan L. Rosner \support}
\Address{\efi}
\vfill
\begin{Abstract}
Quark and lepton flavor physics presents us with a basic question:
Can we understand the pattern of masses and mixings of the known
quarks and leptons, and how do present and proposed measurements
help to advance that goal?  Topics discussed include the
apparent suppression of new flavor-changing effects, the status of
quark and lepton mixing, the implications of new measurements of CP
asymmetries in heavy quark decays, the impications of forthcoming
experiments on the muon's anomalous magnetic moment and its
transitions to an electron, and what we can hope to learn from
electric dipole moments.
\end{Abstract}
\vfill
\begin{Presented}
DPF 2013\\
The Meeting of the American Physical Society\\
Division of Particles and Fields\\
Santa Cruz, California, August 13--17, 2013\\
\end{Presented}
\vfill
\end{titlepage}
\def\thefootnote{\fnsymbol{footnote}}
\setcounter{footnote}{0}

\section{Introduction}

An outstanding puzzle in particle physics is the pattern of quark
and lepton masses and mixings.  Does it point the way to a deeper structure,
or is it governed by random effects?  In Section \ref{sec:qls} we discuss
this pattern, including the status of mixings, the apparent suppression of new
flavor-changing effects, and new measurements of CP violation in heavy quark
decays.

Some measurements to discern the mass and mixing pattern are noted in
Section \ref{sec:meas}.  These include a forthcoming experiment to obtain
a more precise value of the muon anomalous magnetic moment, proposed searches
for various forms of $\mu \to e$ transitions, and searches for electric
dipole moments.

The big unknown player in this question is dark matter, five times as abundant
as the matter we know.  It is evident in the behavior of galaxies, clusters,
large-scale structure, and gravitational lensing.  Trying to guess the pattern
of the known quarks and leptons without accounting for dark matter may be like
trying to guess the structure of the periodic table knowing only Li, Be, and
their relatives.  Some remarks on the dark matter ``elephant in the room'' are
offered in Sec.\ \ref{sec:dm}, while Sec.\ \ref{sec:conc} concludes.

\section{Masses and mixings of quarks and leptons \label{sec:qls}}

\begin{figure}[htb]
\centering
\includegraphics[width=0.8\textwidth]{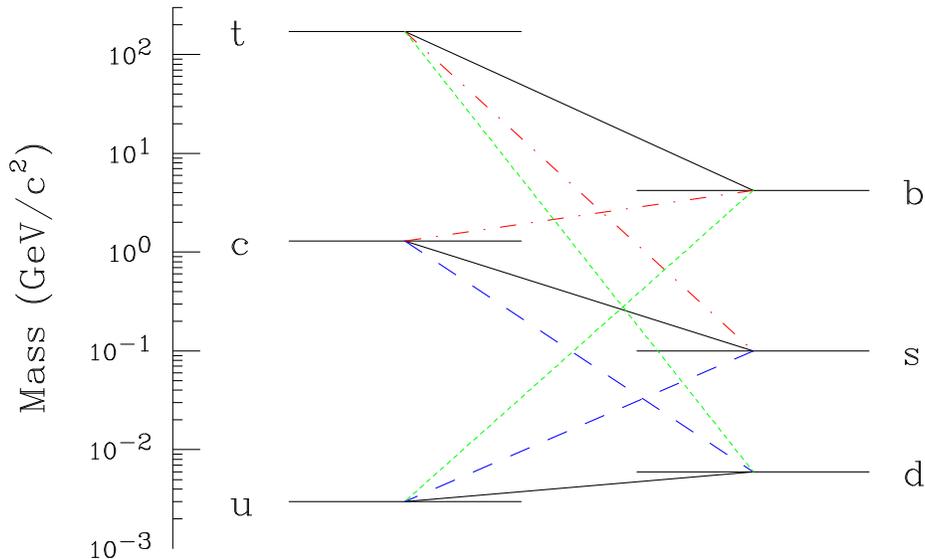}
\caption{The quarks and charge-changing transitions among them.  Relative
magnitudes of transition amplitudes are ${\cal O}(1)$ (black solid lines),
${\cal O}(0.2)$ (blue dashed lines), ${\cal O}(0.04)$ (red dot-dashed
lines), and $< 0.01$ (green dotted lines).}
\label{fig:quarks}
\end{figure}

The Cabibbo-Kobayashi-Maskawa matrix describing charge-changing transitions
among quarks has the hierarchical form \cite{CKMf}
$$
V_{CKM} = \left[ \begin{array}{c c c}
0.974 & 0.225 & 0.0035e^{-i(70^\circ)} \cr
-0.224 & 0.973 & 0.042 \cr
0.0088e^{-i(22^\circ)} & -0.041 & 0.999 \end{array} \right]~,
$$
suggesting that its elements might be correlated with quark masses.  (The
approximate relations $V_{us} \simeq \sqrt{m_d/m_s}$, $V_{cb} \simeq m_s/m_b$
were noted long ago.)  Underlying dynamics might involve logarithms of quark
masses.  In Randall-Sundrum models \cite{Randall:1999ee}, for instance,
fermions could be localized along a fifth dimension, with mixing related to
proximity in this variable.  The mixing pattern is illustrated in Fig.\
\ref{fig:quarks}.

Lepton mixings are a different story, exhibiting a more ``democratic''
pattern \cite{Fogli}:
$$
U_{PMNS} = \left[ \begin{array}{c c c}
0.82 & 0.55 & 0.155 e^{-i \delta} \cr
-0.44 - 0.08 e^{i \delta} & 0.65 - 0.05 e^{i \delta} & 0.61 \cr
0.35 -0.10 e^{i \delta} & -0.52 -0.07 e^{i \delta} & 0.78 \end{array} \right]~.
$$
Aside from the small 13 element, this is not far from
$$
\left[ \begin{array}{c c c}
 2/\sqrt{6} &  1/\sqrt{3} & 0 \cr
-1/\sqrt{6} &  1/\sqrt{3} & 1/\sqrt{2} \cr
 1/\sqrt{6} & -1/\sqrt{3} & 1/\sqrt{2} \end{array} \right]
=
\left[ \begin{array}{c c c}
 0.82 &  0.58 & 0 \cr
-0.41 &  0.58 & 0.71 \cr
 0.41 & -0.58 & 0.71 \end{array} \right]~.
$$
With a sign change in the last row, this is ``tribimaximal''
\cite{Harrison:1994iv} mixing, in which the columns are eigenvectors of a $3
\times 3$ matrix with all 1's.

So, what's the difference between quark and lepton mixings?  The answer could
lie in the seesaw mechanism \cite{seesaw}, a candidate for understanding the
tiny neutrino masses.  The differences between elements of $U_{PMNS}$ and
a tribimaximal $U$ are all less than ${\cal O}(0.1)$ in magnitude, suggesting
that one look for tribimaximal mixing as a first approximation
\cite{Babu:2005ia,McKeen:2007ry}.

Flavor-changing processes are suppressed in the CKM framework.  New fermions,
scalars, and gauge bosons in loops can upset this suppression, but so far
no effects of this sort have been seen.  It appears that whatever new physics
may be hiding in loops, it either respects the CKM pattern (``Minimal Flavor
Violation'' \cite{Isidori:2012ts}) or is associated with a mass scale (e.g.,
$> 10^5$ TeV) far beyond the reach of present accelerators.  As A. Pais used
to say, ``Where's the joke?''

Similar considerations were associated with the attempt to endow the neutral
partner of the Cabibbo current with physical meaning \cite{GIM}.  A charged
current leading to $(d \cos \theta + s \sin \theta) \to u$ would have a neutral
partner in an SU(2) with a flavor-changing part. The inroduction of the charm
quark participating in a transition $(- d \sin \theta + s \cos \theta) \to c$
canceled tree-level flavor-changing neutral currents (FCNC) and led to definite
predictions for loop-level FCNC, e.g., in the transitions $K^+ \to \pi^+ \nu
\bar \nu$ and $K_L \to \pi^0 \nu \bar \nu$.

\section{Present and proposed measurements \label{sec:meas}}

\subsection{Processes related by Minimal Flavor Violation}

In Minimal Flavor Violation (MFV), loop-induced FCNC can generate nonstandard
effects, but they are often correlated with one another.  For example, in MFV
one expects $\Gamma(B_s \to \ell^+ \ell^-)/\Gamma(B_d \to \ell^+ \ell^-) =
|V_{ts}/V_{td}|^2 \simeq 34$.  The Standard Model (SM) predictions for
branching fractions are
$$
{\cal B}(B_s \to \ell^+ \ell^-)=(3.7\pm0.4) \times 10^{-9}~,~~
{\cal B}(B_d \to \ell^+ \ell^-)=(1.1\pm0.15)\times 10^{-10}~.
$$
A combination of results from CMS and LHCb \cite{Bmumu} gives
$$
{\cal B}(B_s \to \ell^+ \ell^-)=(2.9\pm0.7) \times 10^{-9}~,~~
{\cal B}(B_d \to \ell^+ \ell^-)=(3.6^{+1.6}_{-1.4}) \times 10^{-10}~,
$$
consistent with the SM predictions but still leaving room for a deviation
from the ratio predicted by MFV.

The predicted $K^+ \to \pi^+ \nu \bar \nu$ and $K_L \to \pi^0 \nu \bar \nu$
rates are also correlated in MFV \cite{Kronfeld:2013uoa}.  In the SM the
predicted branching fractions are
$$
{\cal B}(K^+ \to \pi^+ \nu \bar \nu) \simeq 8.5 \times 10^{-11}~,~~
{\cal B}(K_L \to \pi^0 \nu \bar \nu) \simeq 2.4 \times 10^{-11}~.
$$

\subsection{CP violation in heavy quark decays}

Observations by CDF \cite{Aaltonen:2011se}, Belle \cite{Ko:2012px}, and LHCb
\cite{Aaij:2011in} suggest that direct CP asymmetries $A_{CP}$ in $D^0 \to K^+
K^-$ and $D^0 \to \pi^+\pi^-$ could be as large as several tenths of a percent.
These asymmetries could be due to non-SM physics, or simply to an enhanced
CP-violating $c \to u$ penguin amplitude \cite{Bhattacharya:2012ah}.  If the
latter, one expects fractional-percent CP asymmetries in other
singly-Cabibbo-suppressed two-body decays such as $D^0 \to \pi^0 \pi^0$ and
$D^+ \to \bar K^0 \pi^+$.  In Ref.\ \cite{Bhattacharya:2013vc} we noted that
such fractional-percent asymmetries can shift the apparent weak phase $\gamma$
extracted from $B \to D K$ decays by up to several degrees.

Large direct CP asymmetries have been reported by the LHCb Collaboration
in certain three-body $B$ decays to charged hadrons \cite{Aaij:2013sfa}.
Even larger asymmtries show up in restricted regions of the Dalitz plot, e.g.,
$$
A_{CP}(B^+ \to \pi^+(\pi^+\pi^-)_{{\rm low}\,m}) = +0.622 \pm 0.075 \pm
0.032  \pm 0.007~,
$$
$$
A_{CP}(B^+ \to \pi^+(K^+K^-)_{{\rm low}\,m})
 =  -0.671 \pm 0.067 \pm 0.028 \pm 0.007~,
$$
where ``low $m$'' refers to low effective mass, defined in \cite{Aaij:2013sfa}.
We have found \cite{Bhattacharya:2013cvn} that these large CP asymmetries can
be interpreted through the interference of SM tree and penguin amplitudes.
Final-state interactions play a crucial role, as do U-spin relations, $K
\bar K$ rescattering, and CPT invariance. The last is important in interpreting
the nearly equal and opposite values of the above two asymmetries.  No new
physics need be invoked.

\subsection{Muon anomalous magnetic moment}

In deep inelastic neutrino scattering, flavor-changing neutral currents are
absent at tree level, but flavor-{\it preserving} effects were crucial in
validating electroweak unification.  One might look elsewhere in
flavor-preserving processes for signs of new physics.  The muon's
anomalous magnetic moment $a_\mu$ is a case in point.  Taking the following
numbers from Ref.\ \cite{HM}, the difference between experiment and theory is
$$
a_\mu({\rm exp}) - a_\mu({\rm th}) = (287)(63)(49) \times 10^{-11}~,
$$
to be compared with\\

\quad Electroweak \cite{Czarnecki:2000id}:~~154(1)(2) $\times 10^{-11}$~;~~
light-by--light:~~70 to 140 $\times 10^{-11}$~;\\
$$
a_\mu^{\rm SUSY} \simeq \pm 130 \times 10^{-11} \left( \frac{100~{\rm GeV}}
{m_{\rm SUSY}} \right)^2 \tan \beta
$$
(see \cite{Czarnecki:2001pv}) which must be larger than the electroweak term
if it is to account for the discrepancy.  Where else do we see such sensitivity
to SUSY?  Flavor-diagonal processes can provide unique windows to new physics.

\subsection{Muon to electron transitions}

In 1962, the muon and electron were seen to be accompanied by separate
neutrinos \cite{Danby:1962nd}.  This explained why the decay $\mu \to e \gamma$
did not proceed with a branching fraction ${\cal B}(\mu \to e \gamma) \simeq
10^{-4}$ \cite{Feinberg:1958zzb}.  Many authors noted the restrictive nature of
the apparent suppression of $\mu \to e$ transitions.  For example, many types
of TeV-scale new physics would be expected to lead to ``rates comparable to or
within a few orders of magnitude of current rate limits \cite{Jungman:1991ru}%
.''

The present situation of $\mu \to e$ transitions has been reviewed in Ref.\
\cite{deGouvea:2013zba}.  Light-neutrino mixing leads to the prediction
$$
{\cal B}(\mu \to e \gamma) = \frac{3 \alpha}{32 \pi} \left |
\sum_{i=2,3} U^*_{\mu i} U_{ei} \frac{\Delta m_{i1}^2}{M_W^2} \right |^2
< 10^{-54}
$$
which is incredibly tiny, but can easily exceed present limits if you
substitute your favorite mixings, $\Delta m^2$, and replace $M_W^2$ by
$\Lambda^2$ where $\Lambda$ is a cutoff.  New physics can induce a dipole
operator (inducing $\mu \to e \gamma$) and/or a four-fermion contact term
($\bar \mu e \bar q q$).  The present upper limit of the conversion rate of
$< 7 \times 10^{-13}$ in Au limits the scale $\Lambda$ to at least $10^3$ TeV
\cite{Bertl:2006up} (quoted in \cite{Bernstein:2013hba}).  Improvement of this
limit to $< 10^{-16}$ in Al will raise the limit on $\Lambda$ by a factor of 7
for this contact term \cite{rhBob}.

\subsection{Electric dipole moments}

Standard Model contributions to electric dipole moments are below the
level of current experiments by several orders of magnitude, but many
scenarios of new physics can give rise to effects observable at present
levels or with various foreseen improvements.  The following values are
taken from Refs.\ \cite{Filippone:2009,Hewett:2012ns}.

If the strong CP phase $\bar \theta$ is zero, CKM contributions to hadron
electric dipole moments need to involve all three quark families and
three-loop Feynman diagrams, leading to a predicted range of $10^{-31}$
to $10^{-32}~e \cdot$cm for the neutron electric dipole moment $d_n$, and
$10^{-33}~e \cdot$cm for $^{199}$Hg.  To generate lepton electric dipole
moments in the SM one needs an additional loop, leading to the predicted
range $10^{-39 \pm 1}~e \cdot$cm for the electron moment $d_e$.

Experimental upper bounds currently are $d_n < 2.9 \times 10^{-26}~e \cdot$cm,
with a factor of 100 improvement foreseen in five years; $d(^{199}{\rm Hg})
< 10^{-27}~e \cdot$cm, with up to $10^5$ improvement claimed possible;
and $d_e < 1.06 \times 10^{-27}~e \cdot$cm (90\% c.l.), where by using cold
molecules (e.g., YbF), a factor of up to $10^4$ improvement may be possible
\cite{Kara:2012ay}.

If they contain CP-violating phases, many models beyond the SM give rise to
observable electric dipole moments.  As an example \cite{McKeen:2012av},
CP violation in the decay of a Higgs boson to $\gamma \gamma$ can lead to
an electric dipole moment through a diagram in which the Higgs boson and
one of the photons interact with a fermion.

\section{The elephant in the room:  dark matter \label{sec:dm}}

\begin{figure}
\begin{center}
\includegraphics[width=0.7\textwidth]{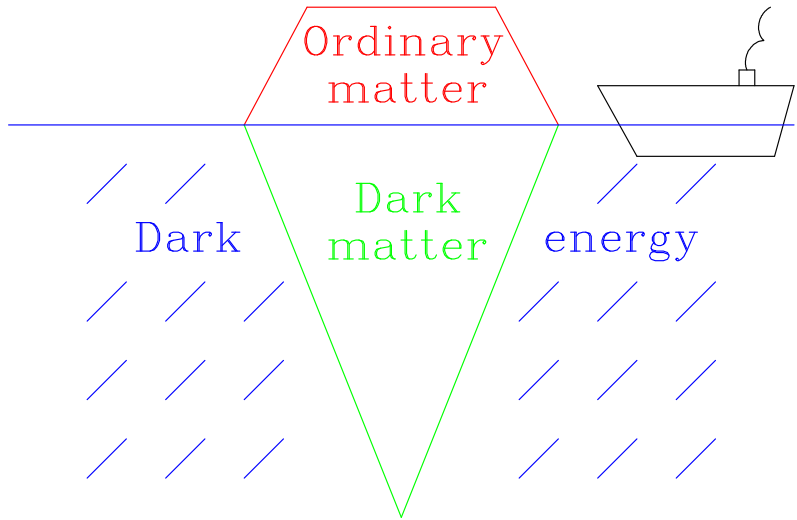}
\end{center}
\caption{Balance between ordinary and dark matter}
\label{fig:berg}
\end{figure}

The preponderance of dark matter (by a factor of five) over ordinary matter
could signify that we are privileged to see only a small subset of gauge
interactions in the SM.  There could exist a ``hidden'' gauge sector $G$ with 
its own exotic charges (see, e.g., Ref.\ \cite{Rosner:2005ec}).  Ordinary
quarks and leptons could be just the tip of a large iceberg (Fig.\
\ref{fig:berg}).

\subsection{Relevance to the flavor problem}

The unseen part of the iceberg could be a clue to the nature of ordinary
matter.  Like blind men, we may be able to put together a coherent
picture of the dark matter ``elephant'' from several pieces of
evidence.  It would be most fortunate, for instance, if some particles
had charges both in the SM and in $G$, as shown in Table \ref{tab:chgs}.

\subsection{Hidden sector and the Higgs boson}

The Higgs boson could be a different tip of the same iceberg.  Its light
mass suggests that the Higgs sector is {\it not} a replay of QCD at a scale
$v/f_\pi \simeq 2650$, where $v = 246$ GeV and $f_\pi = 93$ MeV.  Nonetheless,
composite Higgs theories refuse to die.  A key difference from QCD is that
whereas the lightest $q \bar q$ composites in QCD are pseudoscalar ($J^P =
0^-$), the Higgs boson is at least predominantly a scalar ($0^+$), with upper
bounds on its $0^-$ admixture becoming ever more stringent.  This could be due
to a non-vector-like interaction between constituents of the Higgs boson.

\begin{table}
\caption{Types of matter and their SM and $G$ properties.
\label{tab:chgs}}
\begin{center}
\begin{tabular}{c c c c} \hline \hline
Type of matter & Std.\ Model &    G    & Example(s) \\ \hline
Ordinary       & Non-singlet & Singlet & Quarks, leptons \\
Mixed          & Non-singlet & Non-singlet & Superpartners \\
Hidden         & Singlet     & Non-singlet & $E_8'$ of E$_8 \otimes$
E$_8'$ \\ \hline \hline
\end{tabular}
\end{center}
\end{table}

Some questions for the Higgs and hidden sectors:  (1) If the Higgs boson is
composite, is there one doublet or two?  (2) Do Higgs bosons, quarks, and
leptons share $Q = \pm 1/2$ components \cite{Greenberg:1980ri,%
Fritzsch:1981zh,Fritzsch:1981tt}?  (3) Does a hidden sector play a role in
generating a composite Higgs boson?

\subsection{Two familiar patterns}

\begin{figure}[h]
\includegraphics[width=0.56\textwidth]{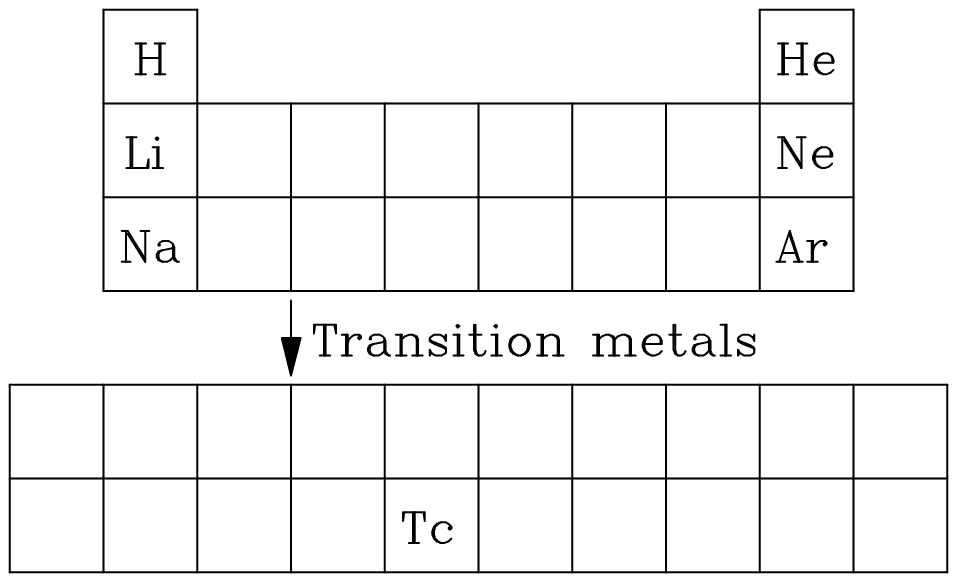} \hskip 0.5in
\includegraphics[width=0.34\textwidth]{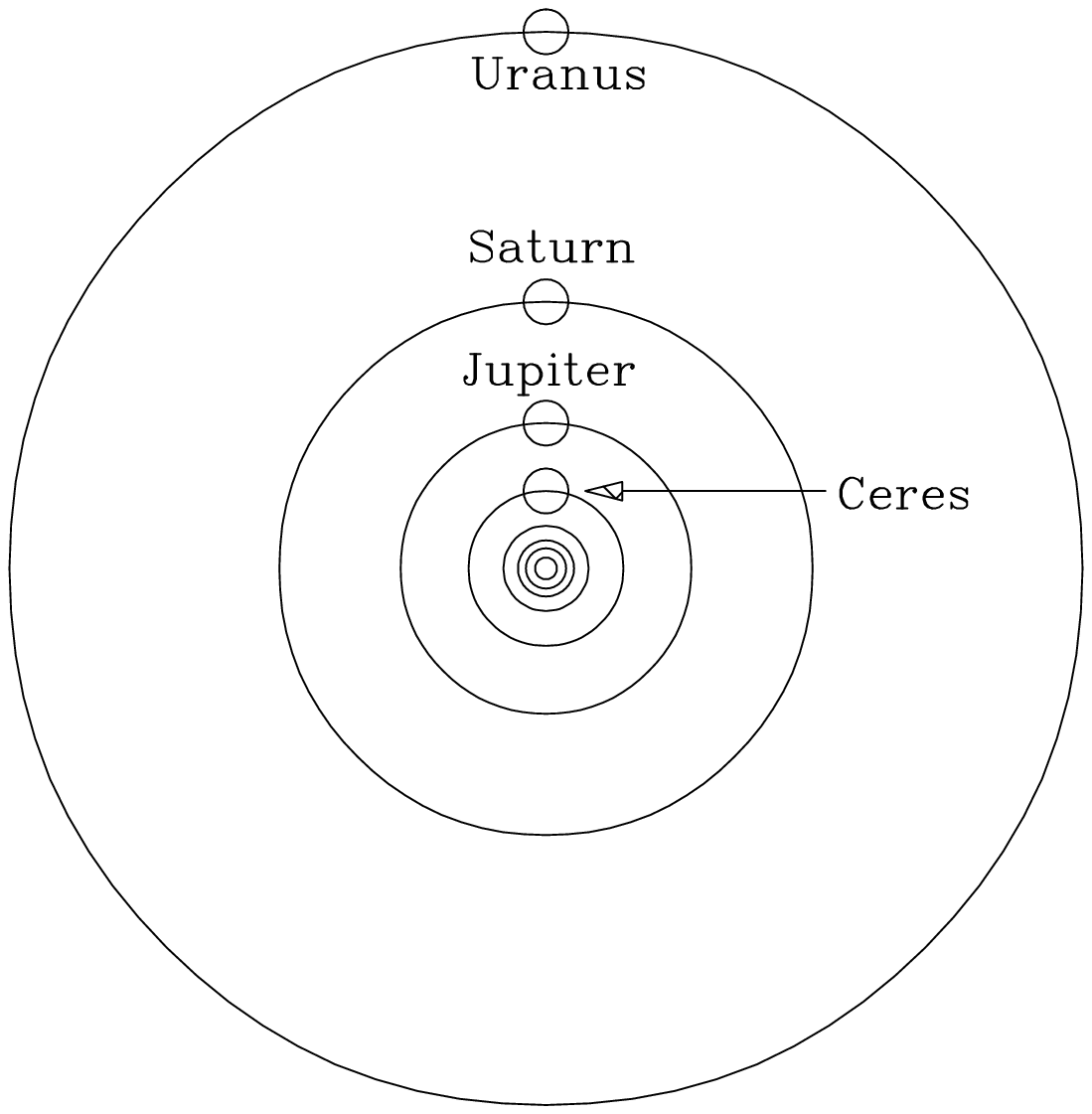}
\caption{Left:  Periodic table of the elements.  Right:  Orbits of inner
planets.}
\label{fig:ptbode}
\end{figure}

In Fig.\ \ref{fig:ptbode} we illustrate two familiar patterns.
The arrangement of the elements is a triumph of quantum mechanics.  Each
element has a different nuclear charge; electron shell structure governs
chemistry.  On the basis of this scheme, the existence of technetium (Tc)
was predicted.  However, imagine if Mendeleev had only the labeled elements
to work with:  he would not have constructed his scheme.  We may be in the
same situation with regard to ordinary and dark matter.

Titius and Bode saw that the planetary orbits followed a simple
regularity, with semi-major axes obeying the law $a ({\rm AU}) = 0.4 +
0.3 k$, where $k = 0, 1, 2, 4, 8, \ldots$.  This law predicted the orbits of
Ceres and Uranus, but failed to predict the orbit of Neptune.  Pluto is
near where Neptune should have been, other dwarf planets don't fit, and there
is no dynamical explanation of the law.  Simulations can give similar
relations, in analogy to the ``anarchy'' pattern in quark and lepton masses
\cite{deGouvea:2012ac}.

\section{Conclusions \label{sec:conc}}

Is the pattern of quark and lepton masses and mixings more like a periodic
table, with an underlying explanation, or like the Titius-Bode description
of planetary orbits, just a step away from a roll of the dice?  So far, we
have no convincing theory.  Further progress awaits better neutrino mixing
measurements (including of the CP phase), improved understanding of the Higgs
sector, and elucidation of the dark sector:  What is hidden from us?

We are in a happy situation I have not seen since the 1960s, when the lack of
a ``Standard Model'' didn't stop us from making progress.  I hope the
situation is about to repeat itelf.

\Acknowledgments

I thank Bob Bernstein, Bhujyo Bhattacharya, and Sheldon Stone for comments on
the manuscript, and Elizabeth Worcester for clarification of the relation
between charged and neutral $K \to \pi \nu \bar \nu$ decays.

\end{document}